\documentclass[groupedaddress,superscriptaddress,aps,showpacs,amsmath,amssymb,floatfix, prb,twocolumn,a4]{revtex4-1}

 \usepackage{graphicx,float}
 \usepackage{subfigure}
\usepackage{bm}
\usepackage{mathrsfs}
\usepackage{txfonts}
\usepackage{CJK}
\usepackage[amssymb]{SIunits}
\usepackage{epsfig}
\usepackage{lipsum}
\usepackage{color}
\usepackage{textcomp}
\usepackage{placeins}

 \newenvironment{SChinese}{%
  \CJKfamily{gbsn}%
 \CJKtilde
  \CJKnospace}{}

\allowdisplaybreaks[2]

\begin{document}

\begin{CJK}{UTF8}{} 
\begin{SChinese}

\title{Generating spin squeezing states and Greenberger-Horne-Zeilinger entanglement using a hybrid phonon-spin ensemble  in diamond}

 \author{Keyu Xia (夏可宇)}  %
 \email{keyu.xia@mq.edu.au}
 \affiliation{ARC Centre for Engineered Quantum Systems, Department of Physics and Astronomy, Macquarie University, NSW 2109, Australia}

 \author{Jason Twamley}
 \affiliation{ARC Centre for Engineered Quantum Systems, Department of Physics and Astronomy, Macquarie University, NSW 2109, Australia}

\date{\today}

\begin{abstract}
  Quantum squeezing and entanglement of spins can be used to improve the sensitivity in quantum metrology. Here we propose a scheme to create collective coupling of an ensemble of spins to mechanical vibrational mode actuated by an external magnetic field. We find an evolution time where the mechanical motion decouples from the spins, and the accumulated  geometric phase yields a squeezing of $5.9~\text{dB}$ for $20$ spins. We also show the creation of a Greenberger-Horne-Zeilinger spin state for $20$ spins with a fidelity of $\sim 0.62$ at cryogenic temperature. The numerical simulations show that the geometric-phase based scheme is mostly immune to thermal mechanical noise.
\end{abstract}



\maketitle

\end{SChinese}
 \end{CJK}

\section{Introduction}
Electronic spins associated with nitrogen-vacancy(NV) centers in diamond posses ultralong coherent times in their ground states at room temperature \cite{LongT2NV1,LongT2NV2}, and can be initialized, controlled and read out using magnetic and optical fields. These features motivate intensive interest in NV-based quantum information processing \cite{NVQIP} and sensing applications. For instance the NV center has been used for quantum computing \cite{NVQComp1,NVQComp2}, cavity quantum electrodynamics system \cite{NVCavityQED}, hybrid quantum interface \cite{PhysRevA.91.042307}, nanoscale magnetometry \cite{NVNanoMagnetometry1,NVNanoMagnetometry2,NVNanoMagnetometry3}, ultrahigh precise solid-magnetometry \cite{PhysRevX.5.041001,PhysRevLett.110.160802,PhysRevA.92.043409}, thermometer \cite{NVMagnetometryThemometer,NVThemometer1,PhysRevB.91.155404}, and nanoscale imaging \cite{NVImaging1,NVSTED} etc.. 

The application of the squeezed spin state (SSS) and the Greenberger-Horne-Zeilinger (GHZ) spin state can boost the precision of quantum metrology \cite{PhysRep.509.89}. Spin squeezing has been typically realized in atomic ensembles \cite{AtomicSpinSqu1,AtomicSpinSqu2,AtomicSpinSqu3,AtomicSpinSqu4}, while the state-of-the-art experiment has achieved $20~\text{dB}$ squeezing using half a million ultracold Rb atoms in a natural atomic trap \cite{AtomicSpinSqu2}. It is of interest to squeeze solid-state spins as this can lead to potentially novel sensor applications. The NV center has a solid-state spin-1 triplet ground state. The squeezing of NV centers has been proposed with the help of strain-induced spin-phonon Tavis-Cummings type interaction \cite{PhysRevLett.110.156402}, however small amounts of thermal excitation can completely inhibit the squeezing. The standard quantum limit in quantum metrology can also be surpassed by using an entangled GHZ state. To date, entangled Bell's states and small GHZ states \cite{GHZ3Qubit,GHZPhoton,SpinNOONstate} have been generated in various systems, but a GHZ spin state with more than $10$ spins has yet to be demonstrated in the literature.

In this paper we describe an approach to engineer the collective coupling of an ensemble of NV centers to a mechanical resonator, mediated by an external magnetic field. Using our protocol, at time $t_m=2m\pi/\omega_m$, where $\omega_m$ is the mechanical oscillation frequency and $m$ is an integer, the mechanical resonator decouples from the NV spin ensemble. We find that the accumulated geometric phase on the NV centers can creates a SSS up to $20$ spins whose squeezing is $\sim 5.9~\text{dB}$ at a large $m$. We also find that our protocol can generate with high fidelity the GHZ state up to $20$ electronic spins. This paper provides a precise numerical investigation of the influence of thermal mechanical noise on the squeezing and the fidelity of the achieved GHZ spin state. 
In contrast to Bennett's work \cite{PhysRevLett.110.156402}, our geometric-phase based scheme is robust against thermal mechanical noise. We note that Zhang et al., very recently, analytically studied the squeezing of $10$ NV centers coupling to mechanical motion \cite{PhysRevA.92.013825}. However, our method is more accurate and can provide a useful prediction for larger spin ensembles. 
Moreover, in contrast to previous work \cite{PhysRevA.92.013825} exploring the trapped nanodiamond, our scheme for spin squeezing and entanglement uses a diamond nanowire. 
We also discuss the generation of GHZ states.

\section{System and model}
\begin{figure}
 \centering
 \includegraphics[width=0.98\linewidth]{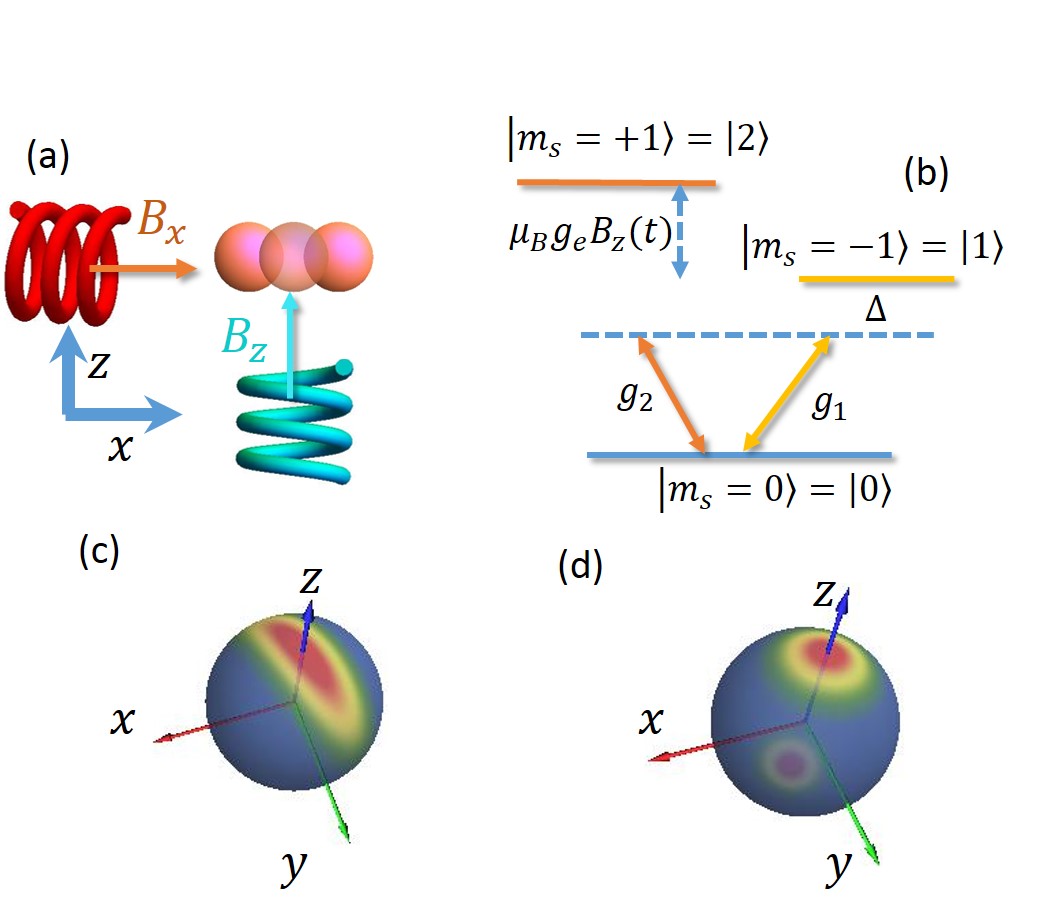} \\
 \caption{(Color online) Schematic diagrams for system squeezing and entangling an ensemble of electron spins in a nanodiamond. (a) An ensemble of NV centers in a nanodiamond oscillates along the $x$-direction with a resonance frequency $\omega_m$. The principal axes of nanodiamond is cut along the $z$-direction. A magnetic field $B_z$ is parallel to $z$ axis, while the magnetic field $B_x(x)$ is applied along the $x$ direction and has a giant gradient along the $x$ direction. (b) The level diagram of a NV centers in a nanodiamond. (c) The Bloch sphere representation of squeezed $N=20$ NV centers. (d) The Bloch sphere representation of the GHZ state of $N=20$ NV centers.} \label{fig:Schematica}
\end{figure}
 The schematic setup for squeezing and entanglement of an ensemble of NV spin centers in a nanodiamond is depicted in Fig. \ref{fig:Schematica}(a).  
 In the setup a nanodiamond oscillates along the $x-$axis with a mechanical frequency $\omega_m$ corresponding to an oscillation period of $T_m=2\pi/\omega_m$. The nanodiamond can be a part of a micro/nano cantilever \cite{DiamondCantilever1}, or attached to a nanotube \cite{DiamondNanotube1}. Alternatively, the nanodiamond can be optically trapped in near vacuum \cite{LevitationNanoDiamond1,LevitationNanoSphere1}. Instead, we choose a single-crystal diamond nanowire with a diameter $d$ and length $L$ for the mechanical resonator \cite{DNT1,DNT2,DNT3,DNT4}.
 We assume the mass of the entire mechanical resonator to be $m$, yielding a zero-point fluctuation of $x_\text{zp}=\sqrt{\hbar/2m\omega_m}$ and a mechanical quality factor of $Q_m$ corresponding to a delay rate of $\gamma_m=\omega_m/Q_m$. The mechanical motion can be quantized as $x=x_\text{zp}(b^\dag + b)$, where $b^\dag$ and $b$ are the creator and annihilation operators of mechanical oscillation. At temperature $T$, the mechanical decoherence due to the thermal excitation is $\bar{n}\gamma_m$ with $\bar{n}= \left[e^{\hbar\omega_m/k_B T}-1\right]^{-1}$, where $k_B$ is the Boltzman constant. Recent experiments \cite{DiamondCantilever1} have demonstrated various diamond micro/nanomechanical resonators with frequencies ranging from $2~\text{\kilo\hertz}$ to several $\text{\mega\hertz}$ and with quality factors $Q_m$ up to $\sim 10^7$. The quality factor, $Q_m$, of a diamond nanowire can also surpass $10^6$ \cite{DNT1}, but the resonance frequency is higher. These high-Q nanomechanical resonators allow us to obtain the required geometric phase with a reasonable magnetic gradient. To make a spin-mechanical hybrid system we can embed NV centers in the end of the diamond nanowire.
 
 We use a magnetic field $\mathbf{B}=B_x(x) \hat{x} + B_z \hat{z}$ to engineer the quantum coupling between the ground states of the NV centers and the mechanical motion, where $\hat{x} (\hat{z})$ is the unit vector along the $x (z)$ direction. Each $\text{NV}$ center has a triplet ground state with levels $|m_s=0,\pm 1\rangle$, as shown in Fig.~\ref{fig:Schematica}(b). The Hamiltonian of the ground states of the $j^\text{\emph{th}}$ $\text{NV}$ center in the magnetic field can be described by  ($\hbar=1$)
 \begin{equation} \label{eq:HSpinB}
  H_j=D_jS_{z,j}^2 + \gamma_B \mathbf{S}\cdot \mathbf{B} \;,
 \end{equation}
 where $D_j\approx 2\pi\times 2.87~\text{\giga\hertz}$ \cite{HamiltonianNV1,HamiltonianNV2}, $\gamma_B\approx 2\pi\times 28 ~\text{\giga\hertz}/\text{\tesla}$ is the gyromagnetic ratio of electron, and $\mathbf{S}$ the electron spin operator. 
 The homogeneous magnetic field component $B_z$ shifts the state $|m_s=-1\rangle=|1\rangle (|m_s=+1\rangle=|2\rangle)$ down (up) by $\gamma_B B_z$, while $|m_s=0\rangle = |0\rangle$ is unshifted, as shown in Fig.\ref{fig:Schematica}(b).
 Here we neglect the hyperfine interaction of which the coupling strengths are typically few \mega\hertz \cite{PhysRevB.85.205203,PhysRevB.79.075203}. It is reasonable because the nuclear spins can be polarized to a selective nuclear spin state and only allow one hyperfine transition \cite{PhysRevB.92.184420,PhysRevB.87.125207,NJP13.025021}.
 We consider the magnetic field $\mathbf{B}$ to possess a giant gradient $G_B=\frac{\partial B_x(x)}{\partial x}$ along the $x$ direction. We engineer this field so that $B_x(x_0)=0$ at the equilibrium position $x_0=0$ of the mechanical resonator. We expand the Hamiltonian in Eq. (\ref{eq:HSpinB}) to the first order of magnetic field gradient that
 \begin{equation}
 \begin{split}
  H_j&=D_jS_{z,j}^2 + \gamma_B B_z (|2\rangle_j\langle 2|-|1\rangle_j\langle 1|)\\
  & + (b^\dag +b) \left[ g_1^{(j)}|1\rangle_j\langle 0|  +g_2^{(j)}  |2\rangle_j\langle 0| +H.c.\right]\;,
  \end{split}
 \end{equation} 
 where $g_1^{(j)} (g_2^{(j)})$ is the magnetic coupling to the transition $|1\rangle (|2\rangle) \leftrightarrow |0\rangle$ of the $j$th spin.
  In this arrangement the interaction between the mechanical motion and the $\text{NV}$ spin ensemble mediated by the magnetic field gradient $G_B$ can be described by the interaction Hamiltonian 
 \begin{equation} 
  H_I^{(x)} = \sum_j  (b^\dag +b)\left[ g_1^{(j)}|1\rangle_j\langle 0|  +g_2^{(j)}  |2\rangle_j\langle 0| +H.c.\right] \;.
 \end{equation} For simplicity, we assume identical coupling that $g_x=g_1^{(j)}=g_2^{(j)}=\gamma_B G_B x_\text{zp}$. Taking $B_z\sim0.1 ~\text{\tesla}$, the frequency $\omega_j=D_j-\gamma_B B_z$ indicating the energy gap between the ground states $|m_s=-1 \rangle(|1\rangle)$ and $|m_s=0\rangle$ is vanishing small but that between $|m_s=1 \rangle (|2\rangle)$ and $|m_s=0\rangle$ can be about $2\pi\times 5.6~\text{\giga\hertz}$. Here we neglect the inhomogeneous broadening in $D_j$ because it is about $1 \text{\kilo\hertz}$, much smaller than the mechanical vibration frequency \cite{LongT2NV1,LongT2NV2}. For the purpose of squeezing and entanglement, at $B_z= 0.1 ~\text{\tesla}$, we can neglect the coupling to the transition between $|2\rangle$ and $|0\rangle$ due to the high energy gap. After dropping the state $|2\rangle$, the Hamiltonian describing the dynamics of the system reduces to
\begin{equation}
  \begin{split}
  H_x =  & \sum_j  \omega_j |1\rangle_j \langle 1| + \omega_m b^\dag b  \; \\
   & + \sum_j g_x (b^\dag +b) (|1\rangle_j\langle 0| + |0\rangle_j\langle 1| ) \;.
  \end{split}
\end{equation}
We set $\omega_j=\omega_0$ and $\Delta = \omega_0 - \omega_m$. We define collective spin operator $J_z=\sum_j (|m_s=-1\rangle_j \langle m_s=-1| - |m_s=0\rangle_j \langle m_s=0|)/2$, $J_+= \sum_j |m_s=-1\rangle_j \langle m_s=0|$ and $J_-= \sum_j |m_s=0\rangle_j \langle m_s=-1|$ to obtain
\begin{equation} \label{eq:Hx}
  H_x =  \omega_0 (J_z + \mathcal{I}/2) +  2g_x  (b^\dag +b) J_x +  \omega_m b^\dag b  \;,
\end{equation}
with $\mathcal{I}=\sum_j (|m_s=-1\rangle_j \langle m_s=-1| + |m_s=0\rangle_j \langle m_s=0|)$, and $J_x=(J_+ + J_-)/2$. We now neglect the term $\omega_0 \mathcal{I}/2$ as it only yields a geometric phase in the evolution. To accommodate the extensively large Hilbert space spanned by a large ensemble of NV centers we apply the Holstein-Primakoff (HP) transformation \cite{HPTransf,DickeQPT1}, yielding $J_z=(a^\dag a - \frac{\mathscr{N}}{2})$, $J_+ = a^\dag \sqrt{\mathscr{N}-a^\dag a}$, and $J_- = \sqrt{\mathscr{N}-a^\dag a}a$ with $\mathscr{N}=N\mathcal{I}$. In the HP picture the Hamiltonian $H_x$ in Eq. (\ref{eq:Hx}) becomes
\begin{equation} \label{eq:HDickex}
 H_\text{HP} = \omega_0 a^\dag a +  \lambda  (b^\dag +b) \bar{J}_x +  \omega_m b^\dag b  \;,
\end{equation}
with $\lambda = 2\sqrt{N}g_x$ and $\bar{J}_x =  (a^\dag \sqrt{\mathcal{I}-a^\dag a/N} + \sqrt{\mathcal{I}-a^\dag a/N}a)/2$ is the Dicke-model collective spin operator.
In the limit $N \rightarrow \infty$ and $\langle a^\dag a\rangle/N \ll 1$, the Hamiltonian $H_\text{HP}$ reduces to
\begin{equation} \label{eq:DickeH}
 H_\text{Dicke} = \omega_0 a^\dag a +  \frac{\lambda}{2}  (b^\dag +b) (a^\dag + a) +  \omega_m b^\dag b \;.
\end{equation}

If we set the magnetic field component $B_z= 0.1 ~\text{\tesla}$ so that one works at the level crossing $\omega_0=0$, the Hamiltonian in the interaction picture of $\omega_m b^\dag b$ becomes 
\begin{equation} \label{eq:Vx}
 V_x = \lambda(b^\dag e^{i\delta t}+be^{-i\delta t}) \bar{J}_x\; ,
\end{equation}
with $\delta=\omega_m$. 
By applying the Magnus' formula \cite{MagnusFormula}, the dynamics for the system can be described exactly, in the absence of decoherence, by the unitary evolution operator 
\begin{equation} \label{eq:Umagnus}
  U_x(t) = e^{iN\theta(t) \bar{J}_x^2} e^{\lambda/\delta\left[\alpha(t) b^\dag - \alpha^*(t) b\right] \bar{J}_x} \;,
\end{equation}
with $\alpha(t) = 1-e^{i\delta t}$ and $\theta(t) = \left(\frac{2g_x}{\delta}\right)^2 (\delta t -\sin\delta t)$. $\theta(t)$ is the unconventional geometric phase which depends only on the global geometric features of operators and is robust again random operation errors \cite{PhysRevLett.90.160402}. $\alpha(t)$ is a periodic function modulating the spin-mechanical coupling. At $t_m=2m\pi/\delta$ for an integer $m$, $\alpha(t)$ vanishes, $\theta(t_m)=2m\pi\left(\frac{2g_x}{\delta}\right)^2$ and the mechanical motion decouples from the NV centers. As a result, the evolution operator for the spin ensemble can be explicitly expressed as 
\begin{equation} \label{eq:Uxt}
 U_x(t_m) = e^{iN\theta(t_m) \bar{J}_x^2} \;.
\end{equation}
Starting from an initial state $|\psi_0\rangle$ the generated state at $t_m$ is $|\psi(t_m)\rangle=U_x(t_m)|\psi_0\rangle$. Note that the squeezing degree of the SSS is only dependent on the available $\theta(t_m)$, which can be controlled with the coupling rate $g_x$ as normal and the number of mechanical period $m$.

Below we will use this evolution operator to squeeze an ensemble of electronic spins for a small $\theta(t_m)=2m\pi\left(\frac{2g_x}{\delta}\right)^2$, see Fig. \ref{fig:Schematica}(c). The scheme can also generate the GHZ state of $N\sim 20$ electron spins if $\theta(t_m)=\pi/2$ is obtainable, see Fig. \ref{fig:Schematica}(d) \cite{PhysRevLett.94.100502,PhysRevA.75.064303}. 
%
%
In contrast to the previous scheme using a Tavis-Cummings type strain-spin interaction in the dispersive regime \cite{PhysRevLett.110.156402}, whose coupling results from a small phononic AC Stark shift, our scheme is superior since the coupling strength is not suppressed by a large detuning. Importantly, our geometric-phase based scheme is immune to many type of noise such as dephasing noise and to first order of spin relaxation since it squeezes the spins via geometric phase control which is known to be quite robust to noise \cite{PhysRevLett.90.160402}. Moreover, the coupling strength $\lambda$ can be tuned by engineering the gradient of the magnetic field. Throughout the further investigation, we will focus on the first cycle of mechanical vibration, i.e. $m=1$ in $t_m$. Later on, we will discuss the operation at a larger integer $m$ requiring a longer evolution time. If the decoherence of the NV centers and the mechanical resonator is small enough, then a large number $m>1$ is preferable because a smaller magnetic field gradient and a larger size of the mechanical resonator are applicable.

The explicit evolution operator in Eq. (\ref{eq:Uxt}) can provide an apparent picture for understanding the unitary evolution of the system. It shows that the initial thermal mechanical occupation is decoupled from spins at $t_1=2\pi/\omega_m$. Here, we set $m=1$ for using the first cycle of mechanical vibration. However it is difficult to include the effect of the mechanical relaxation. The effect may be considerable in some cases due to thermal excitation of the mechanical motion.
To consider the mechanical relaxation, we numerically study the evolution of the combined system by solving the quantum Langevin equation in the Bosonic picture after the Holstein-Primakoff transformation
\begin{equation} \label{eq:QLEq}
 \begin{split}
    \partial \rho/\partial t = & -i [H_\text{HP},\rho] + \mathscr{L}\{(n_\text{th}+1)\gamma_m,b,\rho\}  \\ & +\mathscr{L}\{n_\text{th}\gamma_m,b^\dag,\rho\}\;,
 \end{split}
\end{equation}
where $\mathscr{L}\{\gamma,A,\rho\}= \gamma/2 \{2A\rho A^\dag - A^\dag A\rho - \rho A^\dag A \}$ for $\gamma \in\{n_\text{th}\gamma_m,(n_\text{th}+1)\gamma_\text{m}\}$ and $A\in \{b,b^\dag\}$. Recent experiments have demonstrated ultralong coherence times $T_2\sim 1~\text{\milli\second}$ in ensembles of electron spins in diamond even at room temperature \cite{LongT2NV1,LongT2NV2}. The relaxation time, $T_1$, of NV spin ensemble has also been reported to be much longer, up to several minutes at low temperature \cite{PhysRevLett.108.197601}. Obviously, both $T_1$ and $T_2$ is much longer than our operation time. Moreover, our geometric-phase based scheme is robust again dephasing \cite{PhysRevLett.90.160402}, so we ignore the decoherence of the spin ensemble in the quantum Langevin equation. The noise is dominated by the mechanical thermal noise. As long as $mn_\text{th}\gamma_m/\omega_m \ll 1$, we will achieve a highly squeezed spin state and a high fidelity GHZ state.

In the case where the inhomogeneity of the couplings and the transition frequencies $\omega_j$ are negligible, Dark states are rarely excited. The state of the system can be fully described by the Dicke model defined in Eqs. (\ref{eq:Uxt}) and (\ref{eq:QLEq}). The Dicke state $|J,m\rangle$ with $m\in\{-J, -J+1, \cdots,J+1,J\}$ in the spin picture is equivalent to the Fock state $|J+m\rangle$ in the Bosonic picture \cite{PhysRep.509.89}. Thus, a GHZ state of spin, $|\text{GHZ}_\text{spin}\rangle=(|0\rangle^{\otimes N} + |1\rangle^{\otimes N})/\sqrt{2}$, corresponds to the cat state $|\text{GHZ}_\text{a}\rangle=(|0\rangle + |N\rangle)/\sqrt{2}$ in the Bosonic picture.

In the Bosonic picture, the squeezing degree of spin states $\{|0\rangle, |1\rangle\}$ of NV centers can be evaluated by the squeezing parameter $\xi^2_s$ given by Kitagawa and Ueda as \cite{PhysRep.509.89} $\xi^2_s=1+2\langle a^\dag a\rangle - \frac{2 \langle (a^\dag a)^2\rangle}{N} - 2|\langle\bar{J}_x^2\rangle|$. Correspondingly, the squeezing parameter defined by Wineland et al. is related to $\xi^2_s$ via $\xi^2_R=\left(\frac{N}{2|\langle \vec{J}\rangle|}\right)^2\xi^2_s$ with $|\langle \vec{J}\rangle|=\sqrt{\langle J_x\rangle^2 + \langle J_y\rangle^2 +\langle J_z\rangle^2}$ \cite{PhysRep.509.89}.
While the fidelity of generated GHZ state can be directly calculated as $F=\text{Tr}[\rho |\text{GHZ}_\text{a}\rangle \langle\text{GHZ}_\text{a}|]$ \cite{PhysRep.509.89}.

Before preparation of the squeezing and GHZ states we first optically polarize the $\text{NV}$ centers to their ground states $|m_s=0\rangle$. Then we apply $B_z=0.1~\text{\tesla}$ to bring $\omega_0=0$. Since then, the system begins to evolve under the Hamiltonian Eq. (\ref{eq:HDickex}).

\section{Results}
\subsection{Preparation of squeezing spin states}
Without noise from mechanical motion, the squeezing of the $\text{NV}$ spin ensemble is maximal when $\theta$ is the so-called optimal phase $\theta_\text{opt}=6^{-1/6} (N/2)^{-2/3}$ (see Appendix). The corresponding squeezing parameter defined by Kitagawa and Ueda is $\xi^2_\text{opt}=N^{-2/3}$ \cite{PhysRep.509.89}. $\theta_\text{opt}$ can be very small for a large number of $\text{NV}^-$s. Noting that $\theta=\left(\frac{2g_x}{\omega_m}\right)^2 \omega_m t_m$, we find that the coupling $g_x$ can be quite small in order to squeeze a large ensemble of spins. For a large integer $m$, $g_x$ can be further reduced by a factor of $\sqrt{m}$. This allows the squeezing of an ensemble of spins even for a relative small spin-phonon coupling. A nanomechanical resonator with small mass but high quality factor is preferable because it promises a large coupling under a relative small magnetic gradient and small noise. For our general numerical investigations we choose $Q_m=10^6$ for our mechanical resonator \cite{DiamondCantilever1}. We first investigate the target state at the time $t_1=2\pi/\omega_m$ when the mechanical motion and the spins separate. Then we will look into the target state at a large $m$ which allows a small coupling rate $g_x$ to achieve the optimal phase. The phase $\theta$ can be adjusted at the fixed time $t_m$ by controlling the gradient of the magnetic field. 

\begin{figure}
 \centering
 \includegraphics[width=0.98\linewidth]{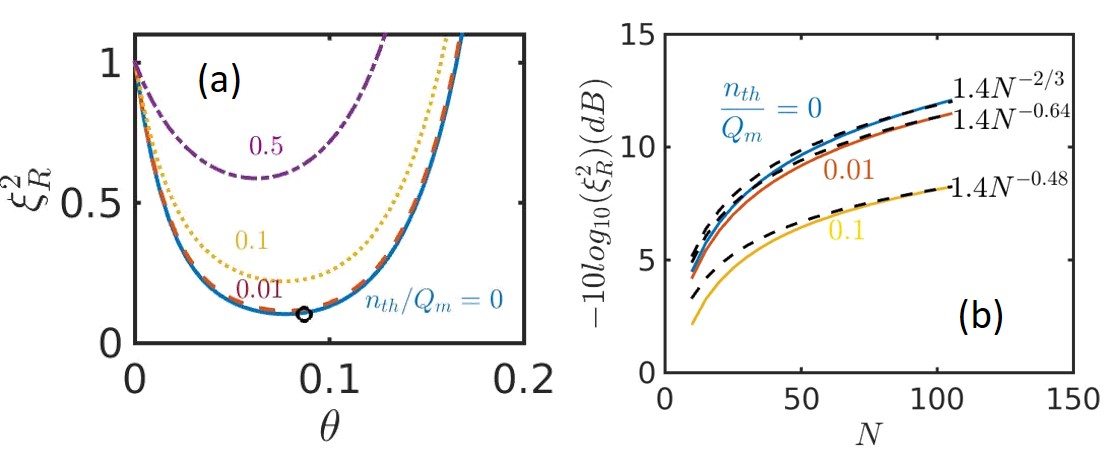}\\
 \caption{(Color online) (a) Squeezing parameter $\xi_R^2$ for $N=50$ spins as a function of $\theta$ for different thermal noise $n_\text{th}/Q_m$. (b) Optimal squeezing parameter $\xi^2_\text{R}~ \text{(dB)}$ as a function of number of spins, $N$ at different thermal noises $n_\text{th}/Q_m=0,0.01,0.1$. The numerical results are fitted by $1.4N^{-2/3},1.4N^{-0.64}$, and $1.4N^{-0.48}$.} \label{fig:squeezing}
\end{figure}

Now we investigate the attainable squeezing by numerically solving the master equation taking into account the mechanical thermal noise.
Firstly, we study the effect of mechanical noise on the squeezing parameter $\xi_R^2$ at the optimal phase $\theta_\text{opt}=0.087~\text{rad}$ at fixed time $t_1$ for $N=50$ spins, see Fig. ~\ref{fig:squeezing}(a). This optimal phase requires $g_x/2\pi=29.4~\text{\kilo\hertz}$ for $N=50$ $\text{NV}$ centers. Such coupling strength is obtainable in a nanomechanical resonator system \cite{NVMR1,PhysRevA.88.033614}. As can be seen in Fig. ~\ref{fig:squeezing}(a), the squeezing parameter first decreases as the phase $\theta$ (or the coupling $g_x$) increases. After reaching the maximal squeezing, it increases as $\theta$ continues to increase. At a low temperature $\bar{n}/Q_m=\bar{n}\gamma_m/\omega_m\leq 0.01$ corresponding to $\bar{n}=10^4$, our scheme can achieve the optimal squeezing $\xi^2_s=N^{-2/3}$ and $\xi^2_R=1.4N^{-2/3}$ at $\theta_\text{opt}$, as shown by the black circle. Even for $\bar{n}/Q_m=0.1$, $\xi^2_R (\xi^2_s)$ only increases to $0.23 (0.15)$ from $0.1 (0.074)$ for $\bar{n}/Q_m=0$. Moreover, the optimal phase decreases very slightly with order variations in $\bar{n}/Q_m$, indicating robust squeezing against mechanical thermal noise. This is an important advantage of this geometric-phase-based squeezing scheme. Even $\bar{n}/Q_m=0.5$ we are still able to squeeze the spins by $\xi_R^2\sim 1.67~\text{dB}$.

By numerically solving the master equation for $N$ up to $105$ we study the dependence of the squeezing on the spin number and the thermal noise. Then we provide an estimate of achievable squeezing for large ensemble of spins, as shown in Fig. ~\ref{fig:squeezing}(b). We calculate the squeezing $\xi^2_R$ for different $N$ at the phase $\theta=\theta_\text{opt}$. When $\bar{n}=0$, the ideal squeezing is $\xi^2_R=1.4N^{-2/3}$, as shown by the blue line and its fitting. The available squeezing slightly reduces to $\xi^2_R=1.4N^{-0.64}$ when the thermal noise dramatically increases to $\bar{n}/Q_m=0.01$, and can still reach $1.4N^{-0.48}$ when the thermal noise reaches $\bar{n}/Q_m=0.1$. Remarkably, the phase uncertainty in measurement, $\Delta\phi=\xi_R/\sqrt{N}$, can be reduced as increasing the number of spins. Note that such squeezing degree of many spins is crucially dependent on the available large gradient,$G_B$, of the magnetic field.

\begin{figure}
 \centering
 \includegraphics[width=0.6\linewidth]{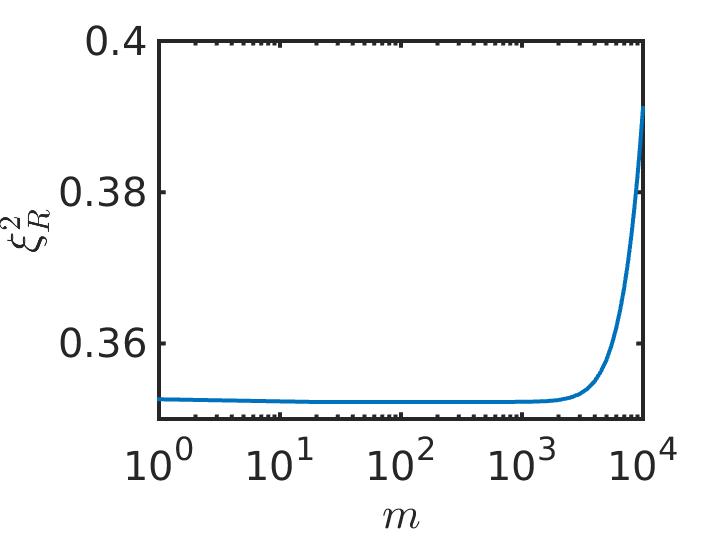}\\
 \caption{(Color online) Squeezing parameter $\xi_R^2$ for $N=10$ spins at different $m$ but fixed $\theta(t_m)$. Here $Q_m=10^6, \bar{n}=10$.} \label{fig:squeezingt}
\end{figure}

As mentioned above, applying a large integer $m$ can relax the required magnetic gradient. As long as $m\bar{n}/Q_m\ll 1$, we can obtain mostly the optimal squeezing. As shown in Fig. \ref{fig:squeezingt} as an example for $N=10$ spins, the squeezing is close to the optimal squeezing $\xi_R^2=0.35$ when $m<2\times 10^{3}$. Even when $m$ increases to $10^4$, we still have $\xi_R^2<0.4$ corresponding to $4 ~dB$ squeezing.

\subsection{Preparation of GHZ states}
In contrast to squeezing only requiring small $\theta$, the generation of GHZ state requires $\theta=\pi/2$ which requires a large coupling strength $g_x$.

\begin{figure}
 \centering
 \includegraphics[width=0.96\linewidth]{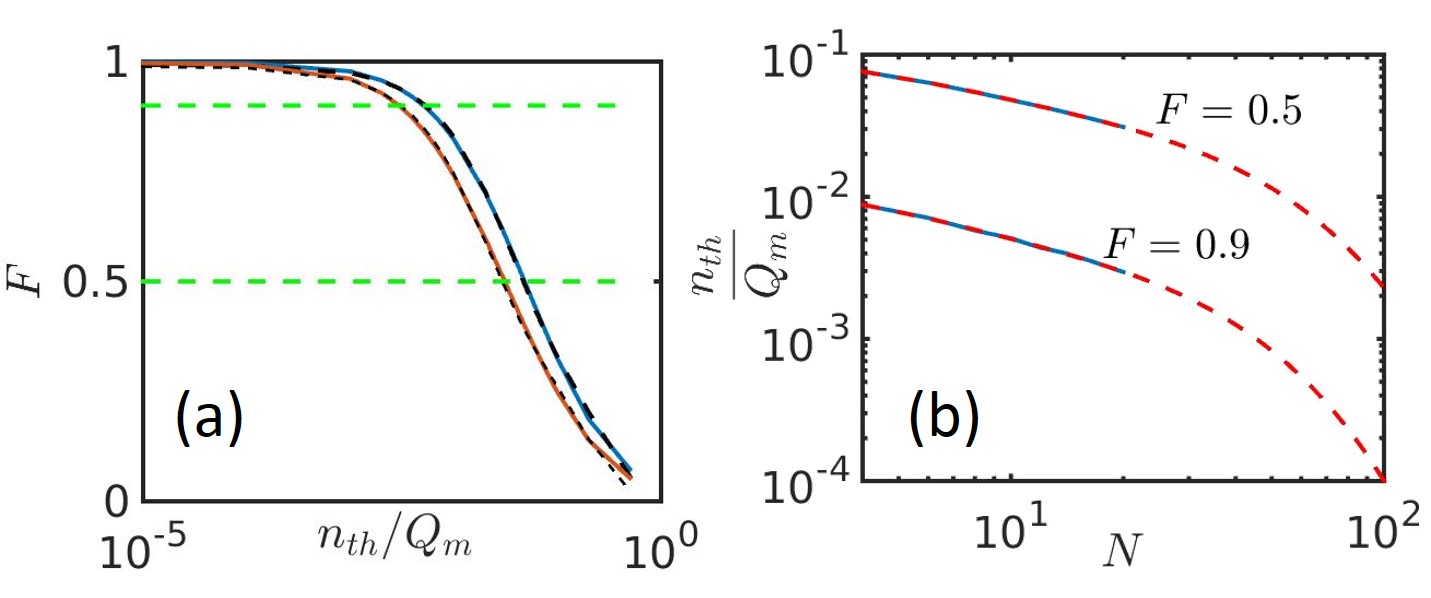}\\
 \caption{(Color online) (a) Fidelity of the generated GHZ state as a function of thermal noise, $n_\text{th}/Q_m$ for $N=10$ (blue line) and $N=20$ (red line) fitted two-term exponential functions, $a e^{b \bar{n}/Q_m} + c e^{d \bar{n}/Q_m}$ with $\{a=0.5208, b=-32.58, c=0.4722, d=-4.241\}$ (dashed black lines) and $\{a=0.5182, b=-54.21, c=0.4704, d=-6.016\}$ (dotted-dashed black lines) respectively. Green lines are guide to eyes at $F=0.5,0.9$. (b) The mechanical thermal decoherence limit for achieving the set fidelity (F=0.5,0.9) for different number of $\text{NV}^{-}$s. Blue lines are numerical results by solving the Master equation. Dashed red lines are the two-term exponential fitting by $a e^{b N} + c e^{d N}$ with $\{a=0.063, b=-0.229, c=0.058, d=-0.032\}$ for $F=0.5$ and $\{a=9.248\time 10^{-3}, b= -0.277, c= 6.816\time 10^{-3}, d=-0.042\}$ for $F=0.9$.} \label{fig:GHZF}
\end{figure}

We first compare the fidelity of the numerically evaluated GHZ state for $N=10$ and $N=20$ $\text{NV}$ centers under mechanical thermal noise, see Fig.\ref{fig:GHZF}(a). When the number of spins is doubled, the fidelity as a function of $\bar{n}/Q_m$ shifts to the left slightly. In both cases the fidelity can be larger than $0.95$ if $\bar{n}/Q_m<10^{-3}$ is available. The fidelity is still higher than $0.5$ as $\bar{n}/Q_m$ decreases to $0.03$ or $0.05$ for $20$ or $10$ spins, respectively. Figure \ref{fig:GHZF}(b) shows the limit of the mechanical thermal noise to achieve a fidelity of $F=0.5$ and $0.9$ for different numbers of spins. It can be clearly seen that we can prepare the GHZ state with a fidelity of $F=0.9$ for up to $20$ $\text{NV}^-$s if $\bar{n}/Q_m<3\times 10^{-3}$ is possible. The fidelity can still be larger than $0.5$ if the thermal noise increases to $\bar{n}/Q_m=2\times 10^{-2}$. Note that this geometric phase protocol can generate GHZ state only for even number of spins.

\begin{figure}
 \centering
 \includegraphics[width=0.6\linewidth]{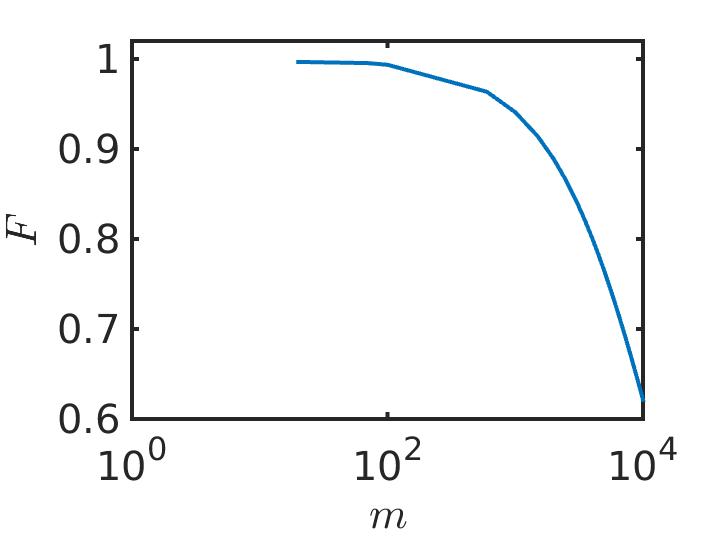}\\
 \caption{(Color online) Fidelity of the GHZ state for $N=10$ spins at different $m$ but a fixed $\theta=\pi/2$. Here $Q_m=10^6, \bar{n}=10$.} \label{fig:GHZt}
\end{figure}

As the generation of the SSS, we can create with a high fidelity the GHZ state at a large integer $m$ as well if $m \bar{n}/Q_m \ll 1$ is met. As shown in Fig. \ref{fig:GHZt}, increasing $m$ will introduce more noise in the target state and subsequently lead to smaller fidelity. However, the fidelity of the GHZ state can be larger than $0.8$ when $m<4\times 10^3$, i.e. $m \bar{n}/Q_m<0.04$.

\section{Magnetic field}
\begin{figure}
 \includegraphics[width=1\linewidth]{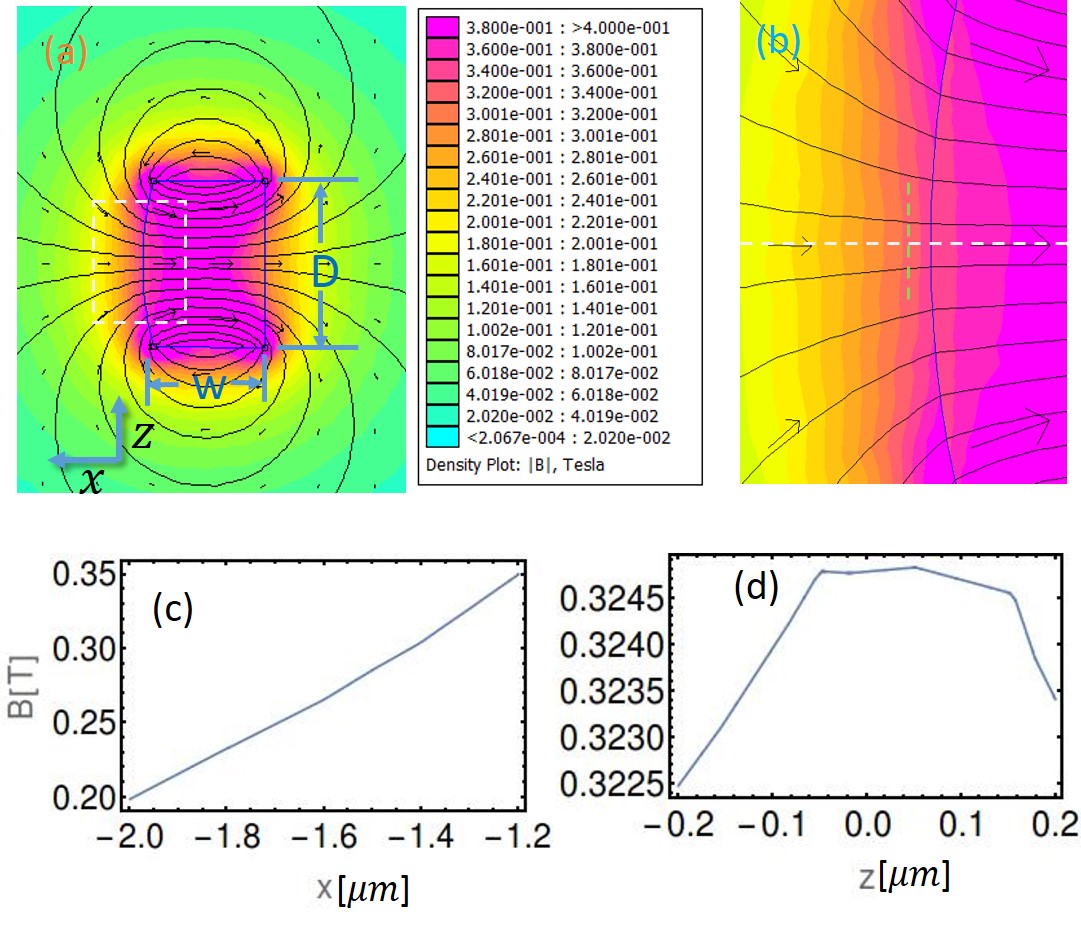} \\
 \caption{(Color online) Magnetic field $B_x$ generated by a two-dimensional permanent magnetic tip made from SmCo27MGOe (simulated with FEMM 4.2). The coordinates of four points are $\{(-1\micro\meter,1.5 \micro\meter), (1\micro\meter,1.5 \micro\meter),(1\micro\meter,-1.5 \micro\meter),(-1\micro\meter,-1.5 \micro\meter),\}$. A $30\degree$ convex arc connects the points $(-1\micro\meter,1.5 \micro\meter)$ and $(-1\micro\meter,-1.5 \micro\meter)$. Therefore, the diameter and the width of the tip is $3~\micro\meter$ and $2~\micro\meter$, respectively. (a) The density and contour plots of the magnetic field surrounding the magnetic tip. The arrows show the direction of the magnetic field. (b) Zoom-in plot of the magnetic field in the dashed white box in (a). (c) The component $B_x$ along the dashed white line in (b). (d) The component $B_x$ along the dashed green line in (b), $100 \nano\meter$ away from the top of the arc.}\label{fig:focuser}
\end{figure}

In this section, we simulate the distribution of the $x$-component of the magnetic field generated by a permanent magnetic tip. We do the two-dimensional simulation with FEMM 4.2. The magnetic field generated by a real three-dimensional magnetic tip is axis symmetrical. We use the material $SmCo27MGOe$ from the material library of FEMM for the magnetic tip. The structure of the tip is shown in Fig. \ref{fig:focuser}(a) (blue box). The tip is  $2 ~\micro\meter$ thick and $3 ~\micro\meter$ wide. The transversal component of the magnetic field ($B_z$) is negligible in the region we are interested in. If we use a plane surface for the tip (see the right hand side of the tip), then the magnetic field in the middle region is weaker than that in edge, and subsequently its transversal distribution is considerably curved. To create a uniform distribution of the field $B_x$, we design a convex curved surface of which the cross section is a $30\degree$ arc connecting the two points, $(-1\micro\meter,1.5 \micro\meter)$ and $(-1\micro\meter,-1.5 \micro\meter)$. As seen in Fig. \ref{fig:focuser}(b) and (d), the transversal distribution of the field $B_x$ is very uniform over about $180~\nano\meter$. Its fluctuation is much smaller than $0.1 ~\milli\tesla$. The gradient of the field $B_x$ is uniform over $500~\nano\meter$ as well, and $\partial B_x/\partial x \approx 2.3\times 10^5~\tesla/\meter$, see Fig. \ref{fig:focuser}(c). Such design can provide us a practical implementation for our proposal. Through simulation we find that the bigger the diameter of the tip is, the smaller the gradient and the wider the transversal uniform region. If we only require a uniform field over $\sim 50~\nano\meter$ in the transversal section, the available gradient can be much larger.

\section{Discussion and conclusion}
Our scheme for squeezing and entangling an ensemble of electronic spins can be realized using a diamond nanowire \cite{DNT1,DNT2,DNT3,DNT4}. 
We consider a single-crystal diamond nanowire with a diameter of $d \sim 9.2~\nano\meter$ and a length of $L\sim 1.45~\micro\meter$ \cite{DNT1}. The existing experimental technology can even make a thiner and lighter DNW to allow a stronger coupling \cite{DNT3,DNT4}. The  mass density of diamond is typically $\rho_m\sim 3000~\kilo\gram/\meter^3$. Thus the mass of the DNW is $\sim 2\times 10^{-19}~\kilo\gram$. The Young's modulus, $E$, of diamond can very from $\sim 40$ to $\sim 900 ~\giga\pascal$ \cite{DNT4,DNT5}. We choose the typical value, $E=300 ~\giga\pascal$, for our DNW yielding a resonance frequency of $\omega_m = 1.88^2\frac{d}{L^2} \sqrt{\frac{E}{16\rho_m}} \approx 2\pi\times 6.1~\mega\hertz$ for its fundamental mode. The zero-point fluctuation is correspondingly $x_\text{zp}\sim 2.2 ~\text{\pico\meter}$. The quality factor of the DNW can be over one million  \cite{DNT1}. So, it is reasonable to take $Q_m=10^6$. At a cryogenic temperature $T= 10~\text{\milli\kelvin}$, we have the thermal excitation of $\bar{n}=33$.

We assume the distance of the closest NV centers along the axis of DNW to be $6~\text{\nano\meter}$ corresponding to a energy shift of about $2\pi\times 0.3~\text{\mega\hertz}$ due to the dipole-dipole interaction \cite{DDI}, which is negligible in comparison with the mechanical resonance frequency. 
The NV centers can be arranged as an array with precise position in an ultrathin film of diamond \cite{NVArrays1,NVArrays2, NVArrays3}. A large magnetic field gradient up to $4\times 10^7~\text{\tesla}/\text{\meter}$ has been reported in the magnetic disk drive system \cite{GradientB1,GiantB2}. In magnetic resonance force microscopy systems, the gradient larger than $10^6~\text{\tesla}/\text{\meter}$ has been realized \cite{GradientB2,GradientB3,GradientB4,GradientB5}. The magnetic field gradient can be nearly constant over $100~\text{\nano\meter}$ in depth \cite{GradientB2}. Below we will apply a gradient about $2\times 10^5~\text{\tesla}/\text{\meter}$ or less.

Now we discuss the experimental implementation for the squeezed spin state and the creation of the GHZ state. We consider $20$ spins embedded in the DNW over $120~\text{\nano\meter}$ along the axis. With choosing a magnetic gradient $G_B=7.5\times 10^4~\text{\tesla}\cdot\text{\meter}^{-1}$. The optimal phase for squeezing, $\theta_\text{opt}=0.16~\text{\rad}$, can be achieved at $t_m=3\times 10^3 T_m$ corresponding to $m n_\text{th}/Q_m \approx 0.1$. The spins can be squeezed by $5.9~\text{dB}$ as a result.
The GHZ state can only be generated at $\theta=\pi/2$. Such large geometric phase requires a large magnetic gradient and a large zero-point fluctuation. Using a magnetic gradient $G_B=2.3\times 10^5~\text{\tesla}\cdot\text{\meter}^{-1}$, we have $g_x/\omega_m\approx 0.0046$ ($g_x/2\pi\approx 28~\text{\kilo\hertz}$) yielding $\theta_x(t_m)=\pi/2$ at $t_{m}=3\times 10^{3} T_m$. 
As a minimum test, lets choose $4$ spins implanted in the DNW requiring a length of at least $ 18~\text{\nano\meter}$ along the axis of the DNW, then we obtain the fidelity of $F=0.62$.
If we can create a uniform magnetic field $B_x$ over $\sim 120~\text{\nano\meter}$ allowing to embed $20$ spins in the DNW, then we can generate the GHZ state with the same fidelity of $F=0.62$ at $t_{m}=3\times 10^{3} T_m$. It is noticeable that, however, the geometric phase protocol can only create the GHZ state for even number of spins. On the other hand, if we are interested in squeezing and entangling a few spins, e.g. $4$ spins, then the applied gradient can be $10^6~\tesla/\meter$ order and the operation can be completely within a few cycles of mechanical vibration. As a result, the squeezing and the fidelity of the GHZ state can be larger.
 
 The evolution period is $2\pi/\omega_m=0.16~\micro\second$ for $\omega_m=2\pi \times 6.1~\mega\hertz$. At $t_m$, the mechanical resonator is decoupled from the spins. So, we need read out the state of spins within time much smaller than $160~\nano\second$.

The large but detrimental Zeeman splitting of the ground states of spins caused by a large magnetic gradient limits the usable number of spins. In the discussion above, we neglect the inhomogeneous broadening of the coupling, $g_x$, caused by the fluctuation of the magnetic field in the transversal direction. It is reasonable only over tens of \nano\meter~ if $G_B=10^6 ~\tesla/\meter$ is applied. Actually, the fluctuation of the magnetic field in the transversal direction will also reduce the fidelity of the target state. These factors limits the obtainable squeezing degree, and the number of entangled spins, and the fidelity as well. By applying smaller gradient $G_B$, the area of uniform field increases. However, the coupling rate reduces. As a result, we need longer evolution time, which is in turn limited by the mechanical quality factor and the thermal excitation.
Therefore, the generation of the SSS state and the GHZ state for $N>20$ spins is quite challenging.

In summary, we couple up to $20$ spins in nanodiamond to a nanomechanical resonator mediated by a magnetic field gradient. At cryogenic temperature, we show a squeezing of $\sim 5.9~\text{dB}$ and the generation of GHZ state with fidelity of $F\sim 0.62$ at the particular time when the mechanical motion decouples from the spins. Our scheme is based on the geometric phase control and therefore is robust against thermal noise, and many other types of noise. 

\section*{Acknowledgement}
This research was supported in part by the ARC Centre
of Excellence in Engineered Quantum Systems (EQuS),
Project No. CE110001013, and in part by the National Nature Science Foundation of China (Grant No. 11204080).

\section*{Appendix A: Optimal phase for maximal squeezing}
\begin{figure}[H]
 \centering
 \includegraphics[width=0.98\linewidth]{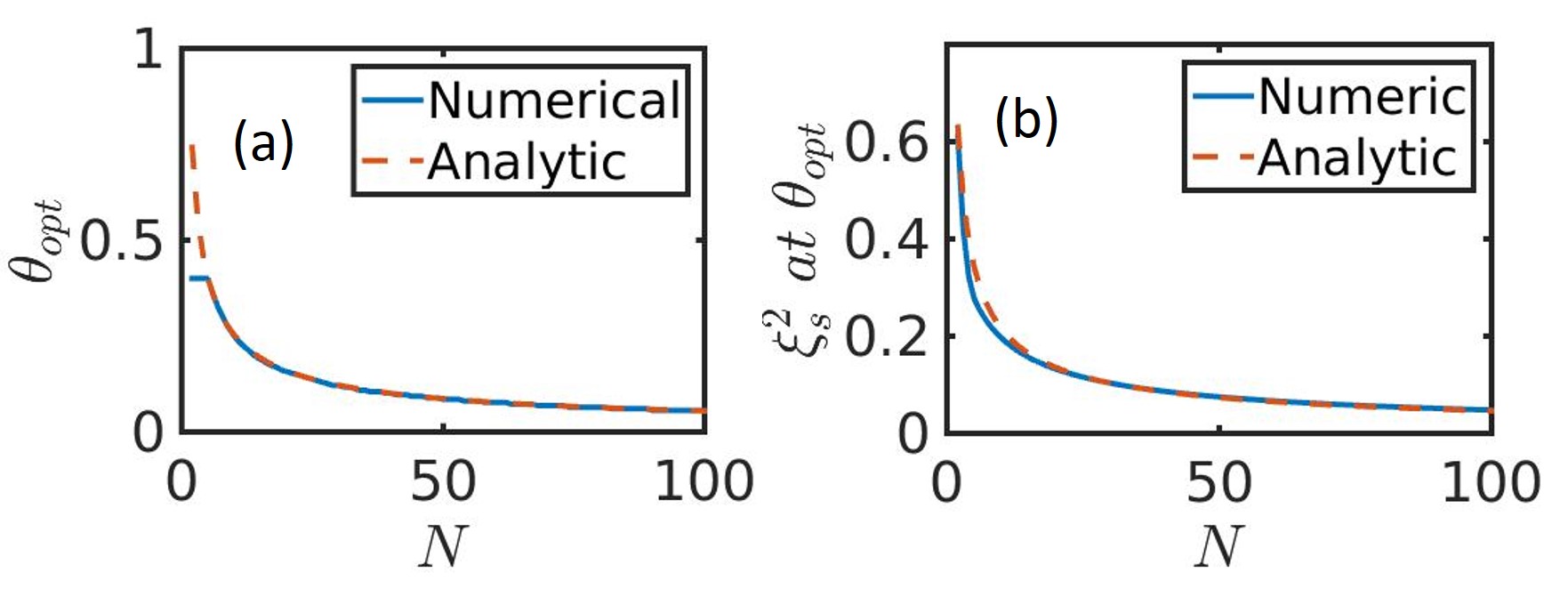}\\
 \caption{(Color online) Checking the optimal phase $\theta_\text{opt}$ for achieving the maximal squeezing in the absence of mechanical decay. (a) The optimal phase $\theta_\text{opt}$ for different spin number $N$. The dashed line is fitted by $\theta_\text{opt}=6^{-1/6}N^{-2/3}$.  (a) The maximal squeezing at $\theta_\text{opt}$ for different spin number $N$. The dashed line is fitted by $\xi_s^2=N^{-2/3}$.} \label{fig:Optimal}
\end{figure}

We numerically study the optimal geometric phase and the corresponding maximal squeezing parameter $\xi_s^2$ as a function of number of spins.
Rather than the formula $\theta_\text{opt}=24^{1/6}(N/2)^{-2/3}$ \cite{PhysRevA.47.5138}, we find the optimal phase in our case to be $\theta_\text{opt}=6^{-1/6}(N/2)^{-2/3}$ by fitting the numerical results, see Fig. \ref{fig:Optimal}.



\end{document}